\documentstyle[preprint, aps]{revtex}

\tightenlines
\newtheorem{theorem}{Theorem}
\newtheorem{acknowledgement}[theorem]{Acknowledgement}

\begin{document}
\title{{\large Calculation of an effective three-ionic interactions potential}\\
{\large in metallic hydrogen}}
\author{E. V. Vasiliu$^{1}$, S. D. Kaim$^{1}$, N. P. Kovalenko$^{2}$}
\address{$^{1}${\it Polytechnic University, 1 Shevchenko av., Odessa, UA-65044,}\\
Ukraine\\
$^{2}${\it State University, 2 Dvoryanskaya str., Odessa, UA-65100, Ukraine}}
\date{\today}
\maketitle

\begin{abstract}
The pair- and three-proton interaction potentials for metal-phase hydrogen
are calculated. Irreducible three-proton interactions are shown to be
essential in the development of the metal hydrogen structure. Possible
manifestations of the three-ion interactions in the structure of quickly
quenched metallic systems are discussed. Existence of amorphous metals is
related to manifestations of the three-ion interactions in non-equilibrium
conditions when amorphisation front travels through a liquid metal.
\end{abstract}

\section{Introduction}

The inclusion of many-particle interactions into consideration of
equilibrium and kinetics properties is the urgent problem in the theory of
condensed matter \cite{1}, \cite{2}. Calculation of the many-particle
interaction potentials is the key to this problem. For simple dielectric and
metal liquids and for crystals pair interaction potentials are known at
least at a phenomenological level,about the three-particle interaction it
cannot be said. This situation is explained by general lack of development
of the theory of highly nonuniform electron gas. Thus to calculate the
three-particle interactions in various condensed systems one needs to know
in an explicit form the three-point nonlinear response functions involving
cooperative phenomena in the many-electron system, which is formed from
atomic electron shells. Nowadays the problems of calculation of such
functions for the majority of systems are not even formulated. It should be
stressed that for the most systems even the polarization operator of
nonuniform electron gas, which makes up these systems, remains unknown. But
there are same exceptions. This refers to the simple metals which can be
closely approximated by the uniform gas of free electrons. A discrete nature
of the ion subsystem can be taken into account by perturbation theory. The
nonlinear response function of uniform gas corresponding to three- and
four-pole diagrams have been calculated within the framework of
many-particle theory of metals \cite{3}.

Calculations of the structural and thermodynamical equilibrium volume
properties of metal-phase hydrogen point up the significance of inclusion of
many-particle ion interactions, and suggest the possibility of a metastable
metallic state at zero external pressure \cite{3}. Comparison of energies
corresponding to different crystal structures have led to the conclusion of
liquidity tendencies in the structure of metal hydrogen. Energy difference
of crystal and liquid phases approaches zero with decreasing pressure \cite
{4}. So the liquid state of metal hydrogen is possible to form at zero
external pressure.

Calculations of energy and other properties of crystal metals are carried
out in the reciprocal space and do not require the explicit form of
interaction potentials of ion clusters. In amorphous, liquid and
heterogeneous metals the calculations in configurational space are
preferable. It provides insight into the nature of local atomic order and
gives the assessment of correlations in the positions of atom clusters.

The occurrence of a maximum in the temperature dependence of the third
virial coefficient evidences directly the existence of three-particle
interactions in simple dielectric liquids \cite{1}. Taking into account the
three-atom interaction in the asymptotic form of the Akselrod-Teller
three-dipole interaction potential requires the inclusion of a cutoff
parameter (the characteristic length) into the theory and does not explain
the wave-vector dependence of the third virial coefficient (on the example
of krypton) \cite{5}.

In metals the many-ion interactions are directly proven to exist by the
anomalies in phonon spectra which could not be reduced to the Kohn's ones,
and conform with the singularities of the many-pole diagrams of uniform
electron gas \cite{3}.

Interaction potential in liquid Na and K have been calculated previously 
\cite{6} for three-ion equiangular configurations. The pair-interaction
potentials for ions in metal hydrogen have been calculated in \cite{7},
where the third-order contributions with respect to electron-ion interaction
were shown to be of fundamental importance for development of attractive
part of potential. Pair interactions in a liquid metal hydrogen have been
calculated within the context of the density-functional method \cite{8}. The
potentials were formulated in terms of direct correlation functions of
electron-ion system.

Here we present the pair- and three-ion interaction potentials in metal
hydrogen at the Wigner-Seitz radius $r_{S}=1,65$ calculated on a basis of
the many-particle theory of metals \cite{3} within the third-order
perturbation theory. The computed array of values of three-ion interaction
potential makes possible the further calculations of the contributions of
these interactions into various properties of metal hydrogen. Possible
outcomes of the three-ion interactions affecting local atomic order in
amorphous metal systems are discussed.

\section{Calculation Results}

In the context of many-particle theory of non-transition metals the energy
of a metal is calculated by the use of the adiabatic approximation for
electron-ion system. The electron gas energy $E_{e}$, provided that ions
positions are fixed, can be calculated with the theory of perturbations in
the potentials of electron-ion and electron-electron interactions \cite{3}.
The energy $E_{e}$\ fulfills the role of an electron term upon the treatment
of the properties of ion subsystem in metal. Together with kinetic energy of
ions and energy of their direct interaction, the energy $E_{e}$ comprises
the effective Hamiltonian of the ion subsystem of metal. $E_{e}$ can be
considered as a sum of contributions independent of ion positions, dependent
on the locations of separate ions, ion pairs, triplets, etc. \cite{3}:

\begin{equation}
E_{e}=\varphi _{0}+\sum_{n}\varphi _{1}({\bf R}_{n})\ +\ \frac{1}{2!}%
\sum_{m\neq {n}}\varphi _{2}({\bf R}_{n},{\bf R}_{m})\ +\ \frac{1}{3!}%
\sum_{m\neq {n}\neq {l}}\varphi _{3}({\bf R}_{n},{\bf R}_{m},{\bf R}_{l})\
+\cdots
\end{equation}

Each term of the series (1) describes interactions of ion groups through the
surrounding electron gas and can be represented as a power series in the
potential of electron-ion interaction \cite{3}:

\[
\varphi _{2}({\bf R}_{1},{\bf R}_{2})\quad =\quad \sum_{i=2}^{\infty }\Phi
_{2}^{(i)}({\bf R}_{1},{\bf R}_{2}), 
\]
\begin{equation}
\varphi _{3}({\bf R}_{1},{\bf R}_{2},{\bf R}_{3})\quad =\quad
\sum_{i=3}^{\infty }\Phi _{3}^{(i)}({\bf R}_{1},{\bf R}_{2},{\bf R}_{3}),
\end{equation}
etc., where $\Phi _{n}^{(k)}({\bf R}_{1},\ldots ,{\bf R}_{n})$ represents
the indirect interaction of the $n$ ions through electron gas in the $k$%
-order perturbation theory in electron-ion interaction.

The indirect interaction of two ions described by the potential $\Phi
_{2}^{(2)}\left( \left| {\bf R}_{1}-{\bf R}_{2}\right| \right) $ is well
studied. The potential $\Phi _{2}^{(3)}\left( \left| {\bf R}_{1}-{\bf R}%
_{2}\right| \right) $, corresponding to the energy of indirect two-ion
interaction in the third-order in potential of electron-ion interaction, has
been calculated previously in \cite{6} for liquid sodium and potassium, and
also for metal hydrogen \cite{7}. It is not difficult to obtain an
expression for the $\Phi _{2}^{(3)}\left( \left| {\bf R}_{1}-{\bf R}%
_{2}\right| \right) $:

\[
\Phi _{2}^{(3)}(R)\quad =\quad \frac{3}{4\pi ^{4}}\int\limits_{0}^{\infty
}dq_{1}q_{1}^{2}\int\limits_{0}^{\infty
}dq_{2}q_{2}^{2}\int\limits_{-1}^{1}dxV(q_{1})V(q_{2})V(q_{3})\times 
\]
\begin{equation}
\times \Gamma ^{(3)}(q_{1},q_{2},q_{3})\frac{\sin (q_{1}R)}{q_{1}R},
\end{equation}
where $V(q)$ is the form-factor of the electron-ion interaction potential; $%
\Gamma ^{(3)}(q_{1},q_{2},q_{3})$ is the sum of three-pole diagrams.

In the third-order perturbation theory in $V(q)$ the indirect three-ion
interaction (the irreducible three-ion interaction) is defined by the
expression \cite{6}:

\[
\Phi _{3}^{(3)}(R_{1},R_{2},R_{3})\quad =\quad \frac{3}{2\pi ^{4}}%
\int\limits_{0}^{\infty }dq_{1}q_{1}^{2}\int\limits_{0}^{\infty
}dq_{2}q_{2}^{2}\times 
\]
\[
\times \int\limits_{-1}^{1}dzV(q_{1})V(q_{2})V(q_{3})\Gamma
^{(3)}(q_{1},q_{2},q_{3})\int\limits_{0}^{1}dx\times 
\]
\[
\times \cos \Biggl(x\biggl(q_{1}R_{1}\frac{R_{1}^{2}+R_{2}^{2}-R_{3}^{2}}{%
2R_{1}R_{2}}+q_{2}R_{2}z\biggr)\Biggr)\times 
\]
\[
\times {J_{0}}\Biggl(q_{1}R_{1}\biggl(1-x^{2}\biggr)^{1/2}\biggl(1-\frac{%
(R_{1}^{2}+R_{2}^{2}-R_{3}^{2})^{2}}{4R_{1}^{2}R_{2}^{2}}\biggr)^{1/2}\Biggr)%
\times 
\]
\begin{equation}
\times {J_{0}\Biggl(q_{2}R_{2}\biggl(1-x^{2}\biggr)^{1/2}\biggl(1-z^{2}%
\biggr)^{1/2}\Biggr)},
\end{equation}
where $J_{0}(x)$ is the Bessel function of zero order; $z\equiv \cos ({\bf q}%
_{1},{\bf q}_{2})$; $q_{3}=(q_{1}^{2}+q_{2}^{2}+2q_{1}q_{2}x)^{1/2}$; $%
R_{1},\ R_{2},\ R_{3}$- the distances between the vertices of a triangle
formed by the protons.

The pair- and three-proton interaction potentials were calculated at the
Wigner-Seitz radius $r_{S}=1,65$ which corresponds to zero pressure in the
zeroth model of a metal. A permittivity function in the Heldart-Vosko form
was employed.

Figure 1 shows computed potentials of the two-proton interactions $%
e^{2}/R+\Phi _{2}^{(2)}(R)$, $\Phi _{2}^{(3)}(R)$, and $\varphi ^{\ast
}(R)=e^{2}/R+\Phi _{2}^{(2)}(R)+\Phi _{2}^{(3)}(R)$. It is obvious that the
interaction $\Phi _{2}^{(3)}(R)$ significantly renormalizes potential $%
e^{2}/R+\Phi _{2}^{(2)}(R)$. Noteworthy is a minimum in the repulsive part
of the potential $\varphi ^{\ast }(R)$, which arises as a result of the
contribution $\Phi _{2}^{(3)}(R)$. At $r_{S}=1,72$ this minimum becomes
deeper and as a result the potential $\varphi ^{\ast }(R)$ takes the form
typical for simple metals. At $r_{S}<1,65$ the minimum in the repulsive part
of $\varphi ^{\ast }(R)$ turns shallow and its position shifts towards
smaller $R$.

The potentials of irreducible three-proton interaction $\Phi
_{3}^{(3)}(R_{1},R_{2},R_{3})$ calculated at different lengths of sides of a
three-proton triangle are listed in Table. It should be noted that the
number of various $(R_{1},R_{2},R_{3})$ sets is restricted by the triangle
axiom to which the distances $R_{1},R_{2}$ and $R_{3}$ are subject.

The calculation results can be presented more conspicuously and
informatively as plots of versus geometry parameters for selected
configurations of protons. Thus, Figure 2 demonstrates the potential curve $%
\Phi _{3}^{(3)}(R,R,R)$, that is, one for proton equiangular triplets. The
potential $\Phi _{3}^{(3)}(R,R,R)$ corresponds to strong attraction of the
triplet at short distances and oscillates at the long ones.

Figure 3 shows plots of the potential $\Phi _{3}^{(3)}$ for isosceles ion
triplets. It is clear that attractability of proton triplet is actually
short-range. An equilibrium local atomic order results from the balance of
direct proton interactions, the indirect two-proton ones, and of the
irreducible three-proton interactions.

Figure 4 presents a proton energy plotted against the distance to the two
others (the isosceles proton triplet). Two protons are placed in abscissa
axis at the points (1,0,0) and (-1,0,0) (distances in atomic units). The
third proton is sited in ordinate axis. In Fig. 4 is shown that inclusion of
the irreducible potential $\Phi _{3}^{(3)}$ is of fundamental importance for
determination of the equilibrium local order in proton spacing. Attractive
nature of the $\Phi _{3}^{(3)}$ potential have to reduce the average
interionic distances in equilibrium state of metal.

\section{Discussion and Conclusion}

A comprehensive analysis of the atomic properties of simple metals leads to
the conclusion that of basic importance is the consideration of three-ion
interactions as in polyvalent metals \cite{9}, and also in metal hydrogen
where the electron-ion interaction is free from a non-coulomb part \cite{3}.
Calculations of the two-ion interaction potentials for various simple metals
display the similarity of their behavior in all such metals \cite{10}. Our
calculations $\Phi _{3}^{(3)}$ for equilateral proton triplets, along with
the results of \cite{6}, demonstrate the similarity of the $\Phi _{3}^{(3)}$
behavior in simple metals.

Results of our calculations suggest certain inferences about the possible
manifestations of three-particle interactions through the structure of metal
systems. The potential $\Phi _{3}^{(3)}(R_{1},R_{2},R_{3})$ is symmetric
about interchange of ions and therefore in equilibrium conditions the most
probable three-ion configurations in liquid metals would be the equilateral
triplets.

Experiments on the quick quenching of metals and alloys show that amorphous
phase is readily obtainable in the polyvalent metals and their alloys \cite
{11} where consideration of many-ion interactions are essential. Under the
quick quenching conditions aluminum alloys form quasicrystalline structures
with specific 5-fold axes of symmetry forbidden for crystals \cite{12}. The
possible reason of quasicrystal formation is manifestation of the three-ion
interactions as the amorphisation front moves through a liquid metal.

When two of the three ions are fixed on a surface or located within
amorphous phase and the third ion remains in the liquid, energetically
advantageous would be an isosceles triplet rather than equilateral one.
Owing to the attractive nature of $\Phi _{3}^{(3)}$ the third ion shifts so
that one of the angles in the triplet will be more than $60^{\circ }$. Such
isosceles triplets can serve as a basis for generation of amorphons in an
amorphous phase. Existence of the 5-fold axes of symmetry in quasicrystals
corresponds to the presence of the isosceles triplets with an angle of $%
72^{\circ }$. Thus one of the manifestation of three-ion interactions can be
the creation of amorphons while propagating the amorphization front in
liquid metal.

\begin{acknowledgement}
The authors are grateful to Prof. Yu. P. Krasny for stimulating discussion.
\end{acknowledgement}

\begin{center}
{\large Table}
\end{center}

\begin{tabular}{|c|c|c|c|}
\hline
$R_{1}$ & $R_{2}$ & $R_{3}$ & $\Phi _{3}^{(3)}(R_{1},R_{2},R_{3})$ \\ 
\hline\hline
0 & 0 & 0 & -0.95758 \\ \hline
1 & 1 & 1 & -0.23684 \\ \hline
1 & 1 & 2 & -0.07582 \\ \hline
1 & 2 & 2 & -0.01523 \\ \hline
1 & 2 & 3 & -0.00326 \\ \hline
1 & 3 & 3 & -0.00353 \\ \hline
1 & 3 & 4 & -0.00411 \\ \hline
1 & 4 & 4 & -0.00167 \\ \hline
1 & 4 & 5 & 0.00130 \\ \hline
1 & 5 & 5 & 0.00116 \\ \hline
1 & 5 & 6 & 0.00005 \\ \hline
1 & 6 & 6 & -0.00055 \\ \hline
2 & 2 & 2 & 0.00117 \\ \hline
2 & 2 & 3 & -0.00032 \\ \hline
2 & 2 & 4 & -0.00283 \\ \hline
2 & 3 & 3 & -0.00184 \\ \hline
2 & 3 & 4 & -0.00161 \\ \hline
2 & 3 & 5 & 0.00029 \\ \hline
2 & 4 & 4 & -0.00028 \\ \hline
2 & 4 & 5 & 0.00065 \\ \hline
2 & 4 & 6 & 0.00010 \\ \hline
2 & 5 & 5 & 0.00038 \\ \hline
2 & 5 & 6 & -0.00024 \\ \hline
\end{tabular}
\qquad \qquad \qquad 
\begin{tabular}{|c|c|c|c|}
\hline
$R_{1}$ & $R_{2}$ & $R_{3}$ & $\Phi _{3}^{(3)}(R_{1},R_{2},R_{3})$ \\ 
\hline\hline
2 & 6 & 6 & -0.00033 \\ \hline
3 & 3 & 3 & -0.00134 \\ \hline
3 & 3 & 4 & -0.00030 \\ \hline
3 & 3 & 5 & 0.00048 \\ \hline
3 & 3 & 6 & 0.00006 \\ \hline
3 & 4 & 4 & 0.00027 \\ \hline
3 & 4 & 5 & 0.00025 \\ \hline
3 & 4 & 6 & -0.00019 \\ \hline
3 & 5 & 5 & -0.00007 \\ \hline
3 & 5 & 6 & -0.00021 \\ \hline
3 & 6 & 6 & -0.00007 \\ \hline
4 & 4 & 4 & 0.00018 \\ \hline
4 & 4 & 5 & -0.00006 \\ \hline
4 & 4 & 6 & -0.00019 \\ \hline
4 & 5 & 5 & -0.00015 \\ \hline
4 & 5 & 6 & -0.00005 \\ \hline
4 & 6 & 6 & 0.00006 \\ \hline
5 & 5 & 5 & -0.00004 \\ \hline
5 & 5 & 6 & 0.00007 \\ \hline
5 & 6 & 6 & 0.00006 \\ \hline
6 & 6 & 6 & 0.00003 \\ \hline
\end{tabular}

(distances in the Table in atomic units, energy in Ry)

\begin{center}
{\Large Figure Captions}
\end{center}

Fig. 1. Potential of the two-proton interactions: 1 - $e^{2}/R+\Phi
_{2}^{(2)}(R)$; 2 - $\Phi _{2}^{(3)}(R)$; \quad \quad \quad \quad 3 - $%
\varphi ^{\ast }(R)=e^{2}/R+\Phi _{2}^{(2)}(R)+\Phi _{2}^{(3)}(R)$

Fig. 2. Potential $\Phi _{3}^{(3)}(R,R,R)$

Fig. 3. Potential $\Phi _{3}^{(3)}$: 1 - $\Phi _{3}^{(3)}(1\,a.u.,R,R)$; 2 - 
$\Phi _{3}^{(3)}(2\,a.u.,R,R)$; 3 - $\Phi _{3}^{(3)}(3\,a.u.,R,R)$

Fig 4. Potential $\varphi (R)=2\varphi ^{\ast }(R)+\Phi
_{3}^{(3)}(2\,a.u.,R,R)$: 1 - $2\varphi ^{\ast }(R)$; 2 - $\Phi
_{3}^{(3)}(2\,a.u.,R,R)$; \quad 3 - $\varphi (R)$ (see explanations in the
text)

\end{document}